\documentclass[%
 reprint,
amsmath,amssymb,
aps,
prb,
]{revtex4-2}

\usepackage{graphicx}
\usepackage{dcolumn}
\usepackage{bm}
\usepackage{ragged2e}
\usepackage[T1]{fontenc} 
\usepackage{lmodern}
\usepackage{xcolor}
\usepackage{amsmath} 
\usepackage{hyperref}
\usepackage[english]{babel}

\renewcommand{\thetable}{\arabic{table}}
\renewcommand{\eqref}[1]{\textbf{(\ref{#1})}}

\begin{document}
\sloppy
\hyphenpenalty=10000
\exhyphenpenalty=10000
\makeatletter
\renewcommand{\fnum@figure}{\textbf{Figure~\thefigure}}
\makeatother

\makeatletter
\renewcommand{\fnum@table}{\textbf{Table~\thetable}}
\makeatother

\preprint{APS/123-QED}


\title{Sliding and stacking thin homo-monolayers from energy landscapes and combinatorics}

\author{Jose D. Mella$^{1,*}$, Jose Luis Cabellos$^2$, Bernardo S. Mendoza$^3$, Salvador Barraza-Lopez$^{1}$}
\email{jmella@uark.edu, sbarraza@uark.edu}
\affiliation{%
1. Department of Physics, University of Arkansas, Fayetteville, Arkansas 72701, USA\\ 
2. Universidad Polit\'ecnica de Tapachula, Tapachula, Chiapas C.P. 30830, Mexico\\
3. Centro de Investigaciones en Optica, A.C., León, Guanajuato C.P. 37150, Mexico}%

\date{\today}

\begin{abstract}
{\centering\textbf{Abstract}\par}
\vspace{6pt}
Structure leads to function: Knowing all possible ways to stack and slide identical monolayers enables the design of functionalities such as nonlinear optical responses and ferroelectricity in layered materials. Besides, mounting experimental evidence points to differing stacking configurations in few-monolayer and bulk samples. To exemplify the core concepts for stacking arbitrary monolayers, ground-state and metastable non-moiré SnSe stacks were first identified using density functional theory and employed to validate a ``divide and conquer'' combinatorial approach for determining the most probable stacking configurations in materials with an arbitrary number of identical monolayers. A small number of point groups accounts for all possible stacks, which explains experimental second-harmonic optical responses. Finite-temperature effects are also studied despite the granularity of available data. This comprehensive framework readily generalizes to provide a predictive description of arbitrary metastable few-homomonolayer stacks.

\end{abstract}

\maketitle


\section*{Introduction}

Two-dimensional (2D) materials host functional phenomena such as structural phase transformations \cite{MehboudiNL,chang2016discovery}, ferroelectricity \cite{chang2016discovery,mehboudi2016structural,Wu2017,yuan2019room,poudel2023creating,liang_resolving_2025}, coupled valley and ferroelectric degrees of freedom \cite{Kai2019,Natcommun2025}, piezoelectricity \cite{hinchet2018piezoelectric,hBNpiezo}, magnetism \cite{Burch,zhang20242d},  strong electronic correlations \cite{macdonald,mit,chunli}, superconductivity \cite{jindal2023coupled,cao2018unconventional}, topology \cite{Engelke,Bennett,SBL}, magnetoelectric multiferroicity \cite{Cromin}, magnetoelasticity \cite{Tian_2016,NatCommunME,John}, and others \cite{yang_observation_2024,zhang_intrinsic_2026}. Stacking and sliding further enhance those properties \cite{chen2025ferroelectricity,MagneticBennett,MagneticCoronado,MagneticJun,li2023two}. Besides, a growing body of experimental literature is pointing towards the fact that materials readily have structures different from their putative bulk phases {\cite{chang2019enhanced,kaloni2019atomic,patel_tuning_2026}}. {Examples include  exfoliated rhombohedral graphene \cite{palinkas_identification_2024}, hexagonal boron nitride grown by chemical vapor deposition \cite{fukamachi_large-area_2023}, transition metal dichalcogenides (TMDC) grown by the same technique \cite{yoo_unraveling_2025}, monochalcogenide (MC) flakes produced by physical vapor deposition \cite{shi2023domain} or vapor liquid–solid growth \cite{sutter_stacking_2022}, and CrI$_3$ grown by chemical vapor transport \cite{jang_direct_2024}. Importantly, such behavior still lacks a formal theory though.}


Motivated by these facts, a general process to classify few-monolayer stacks of the same type of non-magnetic monolayers is described here. This work could shed light on {the stacking and physical properties of the} as-grown samples and on those created by robots. We decided to exemplify it on stacks of layered MC compounds, because metastable stacking configurations of SnS, SnSe, and SnTe have been experimentally realized \cite{chang2019enhanced,sutter2021few,shi2023domain,higashitarumizu2020purely,kaloni2019atomic}. Broken inversion symmetry in some of those configurations induces ferroelectricity  and nonlinear optical responses \cite{chang2019enhanced,kaloni2019atomic,shi2023domain,sutter2021few,higashitarumizu2020purely,mao2023giant}. More recently, a coupling between interlayer sliding and in-plane ferroelectricity was observed in metastable-stacked SnSe \cite{sliding_snse}. The switching voltage in that stack increases with sample thickness, suggesting a correlation with the net electric dipole ($\textbf{P}$) and the number of metastable stacking interfaces between monolayers. Experimental second-harmonic generation (SHG) signals suggest the existence of stable stacking sequences \cite{mao2023giant,moqbel2024giant}. Despite experimental progress, theoretical discussions have addressed only four stacking configurations of their trilayers \cite{li2024first}, and a single stacking possibility for a tetralayer \cite{moqbel2024giant}, clearly leaving the vast majority of possible configurations unexplored.

We present our theoretical framework to investigate all possible stacking configurations and their impact on ferroelectric and nonlinear-optical properties of few-monolayer SnSe is {\em systematic}. Specifically, we (i) predict all stable structures, {(ii) discuss and identify the nonlinear-optical signatures that characterize these phases, and (iii) describe the thermal behavior of these structures.}


The article is organized as follows: structural and energy data obtained from density functional theory (DFT) are first employed to determine possible stacking combinations for bilayers. This information yields the total number of metastable stacks for an arbitrary number of monolayers.

Structural, ferroelectric, and energy properties are analyzed for all possible stacks in trilayer and tetralayer SnSe afterwards, and extrapolated to an arbitrary number of monolayers. A ``divide and conquer'' energy function based on pairwise stacking energies is introduced and validated {at this stage}. The role of symmetry and ferroelectricity in determining equivalent structure is examined. This approach permits calculating stacking probabilities for $N$ identical (\textit{i.e.}, {\em homo-}) monolayers using Boltzmann weights, thereby advancing the analysis from stack identification to stack \textit{likelihood}.

We show that the nonlinear second-order optical response is fully determined by the m and mm2 point groups (PGs) in few-monolayer MC compounds, thus reducing the possible complexity of SHG intensity patterns. Additionally, the second-order nonlinear susceptibility tensor of a trilayer with m PG characterizes the optical response of most SnSe stacks.

{A temperature-dependent total energy model is developed for bilayers, and extended to few-monolayer structures. Boltzmann statistics are applied to determine the most probable stacking configurations, demonstrating that (1) an antipolar (AP) phase is the most stable in the bulk limit; (2) thermal trends can be enunciated despite of the granularity of available data; and (3) metastable structures at zero temperature become preferred at finite temperature.}

\section*{Results and discussion}
\subsection*{Determining all stable bilayer structures}\label{sec:bilayers}

To get started, one must first establish whether a monolayer has multiple degenerate configurations and clarify {\em which one} will be used to study stacking configurations. For a fixed choice and orientation of lattice vectors, MC monolayers have four degenerate ground states (GSs) \cite{RevModPhys.93.011001}: two degeneracies arise upon reflection with respect to the plane defining the monolayer, and two extra degeneracies are related to the orientation of $\textbf{P}$. With that precision in mind, the monolayer structure used as the starting point for stacking SnSe monolayers is the GS explicitly built in Ref.~\cite{poudel2019group}.

The number of metastable stacks in a layered material depends on the number of bilayer minima. For example, a hexagonal TMDC bilayer exhibits both a global and a local minimum in the 2H configuration, as well as two additional local minima when the monolayers are rotated by $60^{\circ}$ (3R phase) \cite{marmolejo2022slippery}. Therefore, it has $D=4$ minima in total. For (magnetic) hexagonal CrI$_3$ bilayers, six minima are identified, with an additional six arising from bilayer structures under a relative rotation of $180^{\circ}$, which yields $D=12$ possible structures \cite{poudel2023creating}. The number of minima in a given bilayer depends on the atomistic interface, and not solely on the lattice type.

\begin{figure*}[tb]
    \centering
    \includegraphics[width=0.96\textwidth]{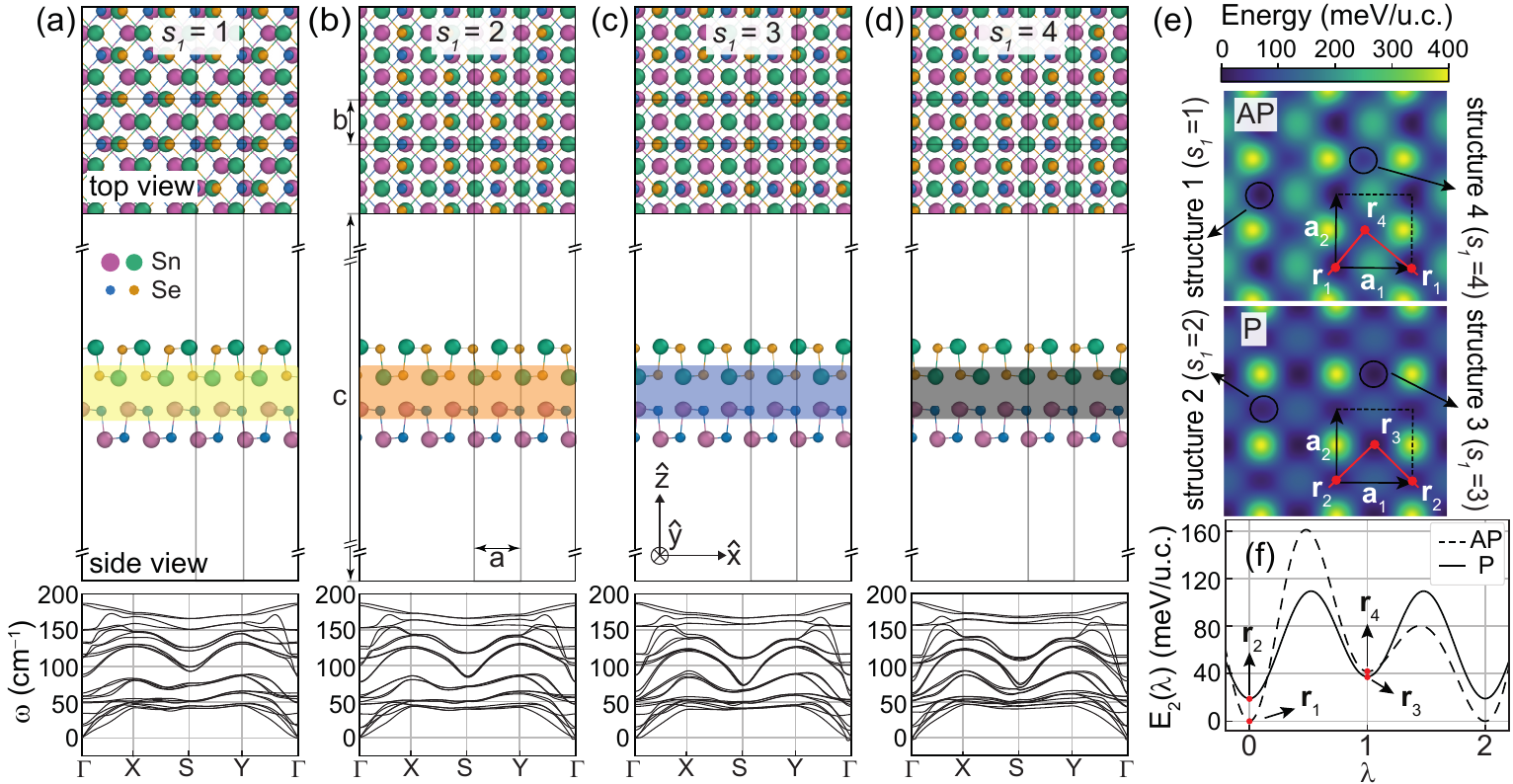}
    \caption{\textbf{Structural, energetic, and vibrational properties of SnSe bilayers.} (a–d): Top and side views of four SnSe bilayers; their structural stability is conveyed by the phonon spectra in the bottom-most panels lacking imaginary frequencies. The two monolayers include background colors for added emphasis, and the lower monolayer remains fixed in all bilayer stacking sequences. The reference Cartesian axes used are in panel (c), where $\hat{x}$ is the armchair direction, $\hat{y}$ the zigzag direction, and $\hat{z}$ the out-of-plane direction. Colors between monolayers indicate the stack interface: $s_1=1$ (yellow), $s_1=2$ (orange), $s_1=3$ (blue), and $s_1=4$ (black) (e) The stability of those bilayers is explained through a zero-temperature energy landscape for SnSe bilayers upon sliding the upper monolayer by $\mathbf{r}$ (and a possible $180^{\circ}$ rotation of that monolayer as well). The upper (lower) panel refers to the bilayer in which the upper monolayer was not rotated (AP) or rotated (P). Minima within the u.c. occur upon a relative in-plane displacement of the upper monolayer by $\mathbf{r}_{1}=\mathbf{0}$, $\mathbf{r}_{2}=\mathbf{0}$, $\mathbf{r}_{3}=(\mathbf{a}_1+\mathbf{a}_2)/2$, or $\mathbf{r}_{4}=(11\mathbf{a}_1+15\mathbf{a}_2)/30$; $\mathbf{a}_1$ and $\mathbf{a}_2$ are lattice vectors. (f) Energy barriers along the paths colored red in panel (e). Jumping between dashed and solid curves requires a macroscopic $180^{\circ}$ rotation of one monolayer, and it is thermodynamically unlikely.}
    \label{fig:1}
\end{figure*}

With the lower monolayer configured as indicated at the start of this section, MC bilayers without a moiré pattern stabilize at either the global energy minimum or at one of the three local minima shown in {\bf Figures \ref{fig:1}(a-d)}. The lower (upper) monolayer is drawn in purple and blue (green and orange) for the Sn and Se atoms, respectively. The colors between monolayers indicate the stack interface: $s_1=1$ (yellow), $s_1=2$ (orange), $s_1=3$ (blue), and $s_1=4$ (black). Their structural stability is ascertained from the lower panels in {\bf Figures \ref{fig:1}(a-d)}, which show vibrational spectra with negligible imaginary modes. {Besides, references \cite{Tian_2016}, \cite{kaloni2019atomic}, \cite{shi2023domain}, \cite{sutter2021few}, \cite{higashitarumizu2020purely},and \cite{mao2023giant} indicate non-bulk stacking sequences, which are built upon those showcased in \bf Figures \ref{fig:1}}

The subscript ``1'' in $s_1$ indicates a stack formed between the first and second monolayers. This notation, though redundant for bilayers, becomes necessary to identify structures having more than two monolayers, where $s_{N-1}$ refers to the stack between monolayers $N-1$ and $N$. This ``positional'' reference is always linked to one of the $D$ stable structures found for bilayers (i.e., $s_{N-1}$ ranges from 1 to $D$).

\begin{table}[tb]
\caption{\textbf{Symmetry, structural and ferroelectric properties of MC bilayers.} LGs and PGs, relative energy $\mathcal{E}_2^{(s_1)}$, lattice parameters $a$ and $b$, interlayer distance $d$, and in-plane $\textbf{P}$ for the MC bilayers shown in {\bf Figures \ref{fig:1}(a-d)}. $P$ is expressed in terms of $P_1\equiv |{\bf{P}}_1|$. Monolayer information in Ref.~\cite{poudel2019group} is included for comparison.}
\label{tab:bilayer}
\centering
\begin{tabular}{ccccccc}
\hline
\hline
\textrm{Structure}& LG, PG & $\mathcal{E}_2^{(s_1)}$ & $a$  & $b$  & $d$  & $P$      \\
             &&($\frac{\text{meV}}{\text{u.c.}}$)&(\AA) &(\AA) &(\AA) & ($P_1$)      \\
\hline
$s_1=1$      & p2$_1$/m11, 2/m&  0.0  & 4.525 & 4.263 & 5.939 & 0 \\
$s_1=2$      & pm2$_1$b, mm2  & 19.0  & 4.487 & 4.279 & 5.983 & $\sim$2 \\
$s_1=3$      & pma2, mm2      & 37.5  & 4.432 & 4.303 & 6.021 & $\sim$2 \\
$s_1=4$      & p2/m11, 2/m    & 41.8  & 4.429 & 4.298 & 6.029 & 0  \\
Monolayer    & pm2$_1$n, mm2  & -       & 4.472 & 4.294 & -     & 1 \\
\hline
\hline
\end{tabular}
\end{table}

The symmetry of the four stable bilayers is characterized next. Structure 1 ($s_1=1$) in {\bf Figure \ref{fig:1}(a)} \cite{deb2017bandgap} belongs to the centrosymmetric layer group (LG) p2$_1$/m11 (\#15) and to PG 2/m. Structure 2 ($s_1=2$), shown in {\bf Figure \ref{fig:1}(b)}, belongs to the non-centrosymmetric LG \cite{itcE} pm2$_1$b ($\#$ 28) and to PG mm2. Structure 3 ($s_1=3$) in {\bf Figure \ref{fig:1}(c)} is non-centrosymmetric and is described by LG pma2 (\#24) and PG mm2. Structure 4 ($s_1=4$) belongs to the centrosymmetric LG p2/m11 ($\#$ 14) and to PG 2/m, and it is displayed in {\bf Figure \ref{fig:1}(d)}. LG and PG classifications are summarized in {\bf Table \ref{tab:bilayer}}.

The DFT energies of all structures with respect to the global minimum are listed in {\bf Table~\ref{tab:bilayer}}, too. The GS corresponds to structure $s_1=1$, whereas the highest-energy structure is $s_1=4$.

The energy landscapes in {\bf Figure \ref{fig:1}(e)} display energy changes resulting from an in-plane displacement $\mathbf{r}$ of the upper monolayer within a (bulk-like) non-rotated AP bilayer configuration ($s_1=1$ or $s_1=4$), or from an in-plane displacement of the upper monolayer on a bilayer configuration following a $180^{\circ}$ rotation of said monolayer (a ``flip''), which turns the bilayer polar (P, $s_1=2$ or $s_1=3$). The in-plane lattice vectors $\mathbf{a}_1=(a,0,0)$ and $\mathbf{a}_2=(0,b,0)$ define the periodicity of the landscape. {\bf Table \ref{tab:bilayer}} lists the structural parameters of bilayers whose energies coincide with these local minima. Interlayer distances $d$ correspond to the $z$-component of the vector joining the geometric centers of each monolayer. Monolayer structural parameters were included in {\bf Table \ref{tab:bilayer}} for comparison.

Energy barriers were estimated by linearly interpolating the lattice and atomistic structure along two discontinuous paths connecting three minima within the AP structure (located at $\mathbf{r}_1$, $\mathbf{r}_4$, and $\mathbf{r}_1+\mathbf{a}_1$) or within the P bilayer (situated at $\mathbf{r}_2$, $\mathbf{r}_3$, and $\mathbf{r}_2+\mathbf{a}_1$). Being specific, the two paths of motion for the upper monolayer in the AP bilayer displayed in {\bf Figures \ref{fig:1}(e)} and {\bf \ref{fig:1}(f)} are ${\bf r}(\lambda)=(1-\lambda){\bf r}_1+\lambda {\bf r}_4$  with $\lambda \in [0,1]$, and  ${\bf r}(\lambda)=(2-\lambda) {\bf r}_4+(\lambda-1)({\bf r}_1+\mathbf{a}_1)$ for the second segment, parameterized with $\lambda$ going from 1 to 2. The energy barriers for the P bilayer are obtained by displacing the upper monolayer along these two paths:
${\bf r}(\lambda)=(1-\lambda){\bf r}_2+\lambda {\bf r}_3$ for $\lambda$ between 0 and 1, and  ${\bf r}(\lambda)=(2-\lambda) {\bf r}_3+(\lambda-1)({\bf r}_2+\mathbf{a}_1)$ for the last segment, for which $\lambda$ varies from 1 to 2. The barriers for transitions between the minima in {\bf Figure \ref{fig:1}(f)} exceed 40 meV per unit cell (u.c.) ({\em i.e.}, 464 K), indicating that once a stable bilayer configuration ``locks-in'' during growth, it may remain so at room temperature. As that transition would require a relative rotation by $180^{\circ}$ between macroscopic monolayers, no thermal pathway exists for transitions between the AP and P configurations.

A key characteristic of SnSe monolayers is their intrinsic and switchable in-plane electric dipole ($\mathbf{P}_1 = 245$ pC/m $\hat{x}$) \cite{RevModPhys.93.011001}, where $\hat{x}$ denotes an in-plane unit vector parallel to the armchair direction. In P bilayers, the electric dipole ${\bf{P}}_2$ increases. This is confirmed in structure 2, where 559 pC/m $\hat{x}$ = 2.28${\bf{P}}_{1} \sim 2${\bf{P}}$_{1}$. It is not exactly twice that of the monolayer, since interlayer interaction increases the lattice parameter, which enhances ${\bf{P}}$ \cite{poudel2019group}. On the contrary, ${\bf{P}}_2=\bf{0}$ in AP bilayers. All possible electric dipole configurations for bilayers are listed in {\bf Table \ref{tab:bilayer}}, and details of the calculation are provided in {\bf Methods}.

{Neighboring bilayers are constrained by the energy landscape ({\bf Figure~\ref{fig:1}(e)}). They are free to explore it at finite temperature, and their relative positions fluctuate within a given energy valley: pairs of monolayers remain trapped on that valley up to the paraelectric transition temperature $T_c$, at which the landscape changes.}

\subsection*{Determining stable configurations of few-monolayer stacks}\label{sec:number}

Having described all possible stable bilayers, the analysis turns to stacks with more monolayers now. The large number of combinations requires setting unique labels: stacking sequences are denoted using one of the four integer labels per interface found in {\bf Figures~\ref{fig:1}(a-d)}. For instance, a trilayer configuration labeled (1,3) indicates that the bottom and middle monolayers form a bilayer consistent with $s_1=1$ ({\bf Figure \ref{fig:1}(a)}), while the middle and top monolayers are stacked as $s_2=3$ ({\bf Figure \ref{fig:1}(d)}).

The number of bilayer minima enables estimation of the total number of metastable stacks. In a stack containing $N$ monolayers, there are $N-1$ interfaces between adjacent monolayers and $D$ bilayer minima. When the first monolayer is fixed, the total number of different stacks for $N \geq 2$ monolayers is given by:
\begin{eqnarray}\label{eq:comb}
   N_{\text{stack}}(D,N)=D^{N-1}.
\end{eqnarray}
{\bf Equation~\eqref{eq:comb}} counts all possible metastable structures, but some configurations may in fact be equivalent. The energetic, symmetry, and ferroelectric properties of trilayers and tetralayers are examined to elucidate equivalent structures next.

\subsection*{Stable trilayer and tetralayer stacks}\label{sec:trilayerTetralayer}

According to {\bf Equation~\eqref{eq:comb}}, there are $4^2 = 16$ MC trilayer configurations, while the number of stable configurations containing four MC monolayers is $4^3=64$. These numbers are vastly larger than the MC stacking arrangements studied previously \cite{li2024first,moqbel2024giant}, showcasing the exhaustive nature of the present study and its predictive power.

The symmetries of trilayer configurations are listed in {\bf Table~\ref{tab:trilayer}}. Four trilayer structures, two AP structures (1,1) and (4,4), as well as structures (2,2) and (3,3) with identical stacking arrangements, are classified under LG pm2$_1$n (\#32) and PG mm2. The remaining twelve trilayer stacking configurations belong to the p11m (\#4) LG and the m PG. Surprisingly, only two LGs and PGs describe all stable trilayer stacks despite the large number of stacking possibilities (only the (4,4), (1,1), (1,4), and (1,2) trilayers have been theoretically studied before \cite{li2024first}). Stick-and-ball trilayer structures are presented in {\bf Supplementary Note S1} and {\bf Supplementary Figure S1}.

Tetralayer AP stacks (1,1,1), (1,4,1), (4,1,4), and (4,4,4) are centrosymmetric, as are structures (2,1,2), (2,4,2), (3,1,3), and (3,4,3). All belong to the 2/m PG. Eight stacking sequences (1,2,1), (1,3,1), (2,2,2), (2,3,2), (3,2,3), (3,3,3), (4,2,4), and (4,3,4) are non-centrosymmetric (mm2 PG). The remaining forty-eight configurations all belong to the p11m LG and m PG. This is what we mean when we say that few LGs and PGs suffice to describe a vast number of stacks. Symmetries are summarized in {\bf Supplementary Tables S1} and {\bf S2}.

The structural energies calculated using DFT in trilayers are reported in {\bf Table~\ref{tab:trilayer}}. All energies are relative to the trilayer GS, which corresponds to the (1,1) stacking configuration. Energy degeneracy occurs for stacking arrangements related by reversal of the stacking sequence, such as (1,2) and (2,1).

The energies relative to the tetralayer GS are also listed in {\bf Supplementary Tables S1} and {\bf S2}. The GS corresponds to the stack (1,1,1). Degeneracies similar to those observed in trilayers are also found in tetralayers (e.g., stacks (1,3,3) and (3,3,1) are degenerate).

{\bf Table~\ref{tab:trilayer}} and {\bf Supplementary Tables S1} and {\bf S2} contain the magnitude of the in-plane net $\bf{P}$ for all trilayer and tetralayer configurations, too. These values are approximated by adding or subtracting the contributions from each monolayer, depending on whether the stacking is P or AP. Among trilayers, four configurations exhibit a polarization of 3$P_1$, while the remaining twelve configurations are approximately equal to $P_1$. The possible magnitudes are 0, $\sim$2$P_1$, and $\sim$4$P_1$ for tetralayers.

The last three columns in {\bf Table~\ref{tab:trilayer}} and {\bf Supplementary Tables S1} and {\bf S2} show the energies, a rescaling parameter, and the parametrized energies derived from the ``divide and conquer’’ approximation. These will be further discussed in the next section.

\begin{table}
\caption{\textbf{Symmetries, energies, electric dipoles, and ``divide and conquer'' approach for trilayer SnSe stacks}. $\mathcal{E}_3(s_1,s_2)$ is the energy obtained from {\bf Equation~\eqref{eq:EN}}.  $\lambda_3=(\mathcal{E}_3^{\text{DFT}}-\mathcal{E}(s_1,s_2))/\mathcal{E}_3^{\text{DFT}}$, whose average is $\bar{\lambda}_3=0.12 \pm 0.06 $. $\mathcal{E}'_3(s_1,s_2)$ is a rescaled energy defined in {\bf Equation~\eqref{eq:renormalized}}. Structures in bold font on the first column are the subset counted in {\bf Equation~\eqref{eq:comb_reduced}}.}
\centering
\label{tab:trilayer}
\begin{tabular}{ccccccc}
        \hline
        \hline
        Str. & LG, PG& $\mathcal{E}_3^{\text{DFT}}$   &$P$& $\mathcal{E}_3(s_1,s_2)$  & $\lambda_3$ & $\mathcal{E}'_3(s_1,s_2)$  \\
    ($s_1,s_2$) & & ($\frac{\text{meV}}{\text{u.c.}}$)          &($P_1$)& ($\frac{\text{meV}}{\text{u.c.}}$)          &             & ($\frac{\text{meV}}{\text{u.c.}}$)            \\
         \hline
     {\bf (1,1)} & pm2$_1$n, mm2&  0.0 & $\sim$1&  0.0  & 0.00 & 0.0   \\
     {\bf (1,2)} & p11m, m      & 23.6 & $\sim$1& 19.1  & 0.19 & 21.4  \\
     {\bf (1,3)} & p11m, m      & 46.3 & $\sim$1& 37.5  & 0.19 & 42.0  \\
     {\bf (1,4)} & p11m, m      & 52.0 & $\sim$1& 41.8  & 0.20 & 46.9  \\
     (2,1)       & p11m, m      & 23.6 & $\sim$1& 19.1  & 0.19 & 21.4  \\
     {\bf (2,2)} & pm2$_1$n, mm2& 42.6 & $\sim$3& 38.3  & 0.10 & 42.9  \\
     {\bf (2,3)} & p11m, m      & 64.3 & $\sim$3& 56.6  & 0.12 & 63.5  \\
     {\bf (2,4)} & p11m, m      & 68.1 & $\sim$1& 61.0  & 0.10 & 68.3  \\
     (3,1)       & p11m, m      & 46.3 & $\sim$1& 37.5  & 0.19 & 42.0  \\
     (3,2)       & p11m, m      & 64.3 & $\sim$3& 56.7  & 0.12 & 63.5  \\
     {\bf (3,3)} &pm2$_1$n, mm2 & 79.6 & $\sim$3& 75.0  & 0.06 & 84.1  \\
     {\bf (3,4)} & p11m, m      & 84.6 & $\sim$1& 79.4  & 0.06 & 88.9  \\
     (4,1)       & p11m, m      & 52.1 & $\sim$1& 41.8  & 0.20 & 46.9  \\
     (4,2)       & p11m, m      & 68.1 & $\sim$1& 61.0  & 0.10 & 68.3  \\
     (4,3)       & p11m, m      & 84.6 & $\sim$1& 79.4  & 0.06 & 88.9  \\
     {\bf (4,4)} & pm2$_1$n, mm2& 87.9 & $\sim$1& 83.7  & 0.05 & 93.7  \\
     \hline
     \hline
\end{tabular}
\end{table}

The symmetry, energetic, and ferroelectric properties identified for trilayers and tetralayers are expected to extend to structures with an arbitrary number of monolayers. In terms of symmetry, the PGs alternate between m and mm2 for odd numbers of monolayers, and among m, mm2, and 2/m for even numbers. Furthermore, $\textbf{P}$ is an even number of $\mathbf{P}_1$ for even $N$ of monolayers, and an odd number times $\mathbf{P}_1$ for odd $N$. {The presence of a proximal substrate lowers structural symmetries; see {\bf Supplementary Table S3}.}

\subsection*{Relative energy of N-monolayer stacks and ``divide and conquer'' estimate}\label{sec:generalization}

We next ponder whether the total stacking energy can be reasonably approximated as a sum of pairwise bilayer energy contributions from the four configurations listed in {\bf Table~\ref{tab:bilayer}}. If such insight is accurate enough, the energy of thicker samples can be estimated without expensive \textit{ab-initio} calculations. Note, at the start, that layered MC compounds change lattice parameters upon stacking ({\bf Table~\ref{tab:bilayer}}). This is due to their metavalent bonding, and has been documented before \cite{poudel2019group}. Here, we assume the stacked monolayers have identical lattice parameters and separations among monolayers (this assumption will be more realistic on more ``traditional'' layered materials like graphite, hBN, or TMDC).

The ``divide and conquer'' energy of a stack containing $N$ monolayers is then defined as follows:
\begin{equation}\label{eq:EN}
\mathcal{E}_{N}(s_1,s_2,\dots,s_{N-1})= \sum_{j=1}^{N-1}\mathcal{E}_{2}^{(s_{j})},
\end{equation}
where $s_{j} \in \{1,2,...,D\}$ is one of the $D$ pairwise stackings for the interface among monolayers $j$ and $j+1$ and $\mathcal{E}_2^{(s_j)}$ its corresponding energy from {\bf Table~\ref{tab:bilayer}}.

The energy obtained for trilayers in {\bf Table~\ref{tab:trilayer}} and for tetralayers in {\bf Supplementary Tables S1} and {\bf S2} exceeds {\bf Equation~\eqref{eq:EN}} by 0–20\%. The difference between DFT values and those obtained from pairwise addition yields average rescaling factors of $\bar{\lambda}_3=0.12\pm0.06$ and $\bar{\lambda}_4=0.16\pm0.06$ for trilayer and tetralayer stacks, respectively. These discrepancies likely arise from charge redistribution in the full stack that is not captured by pairwise sums, as well as from additional structural changes in lattice parameters and interlayer distances. This factor is linearly extrapolated to an arbitrary number of monolayers based on the values obtained for trilayers and tetralayers, resulting in $\bar{\lambda}_N = 0.08 + 0.04(N-2)$ for $N > 3$, and it provides a readily available estimate. We therefore employ rescaled pairwise sums:
\begin{equation}\label{eq:renormalized}
\mathcal{E}'_N(s_1,s_2,...,s_{N-1})\equiv \mathcal{E}_N(s_1,s_2,...,s_{N-1})(1+\bar{\lambda}_N),
\end{equation}
to study larger stacks statistically. The counting degeneracy for structures connected by a reversal of the stacking sequence arising from {\bf Equation~\eqref{eq:comb}} is preserved in this approximation.

{Experimental analysis of a stack containing 8 monolayers yielded an average displacement of approximately 0.05 $\AA$ between adjacent monolayers at room temperature (see Figure 3 in \cite{sliding_snse}). Since this displacement is small compared with the in-plane lattice parameters, it is expected to introduce only a minor correction to the local interface energies.}

\subsection*{Equivalence of metastable stacks}\label{sec:metastable}

{\bf Equation~\eqref{eq:renormalized}} enables statistical analysis of equivalent structures for an arbitrary number of monolayers $N$. We explore the equivalence of energy-degenerate structures connected by reversal of the stacking arrangement in what follows.

\begin{figure}[tb]
    \centering
\includegraphics[width=0.9\columnwidth]{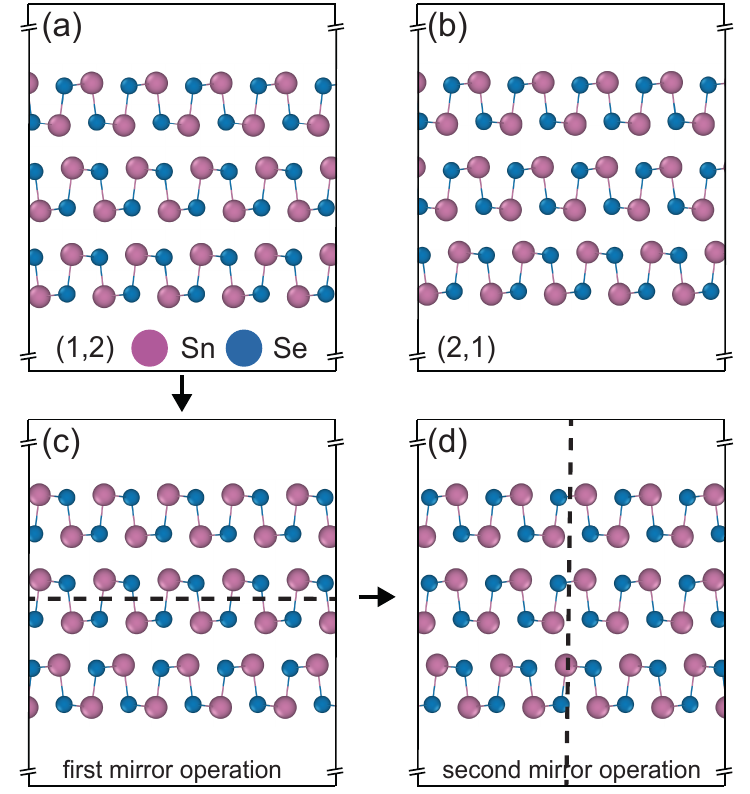}
\caption{\textbf{Equivalence of (1,2) and (2,1) stacks.} Trilayer stacks (1,2) (a) and (2,1) (b). (c) Stack (a) after a horizontal mirror operation (dashed line). (d) Stack (c) after an additional mirror operation perpendicular to the armchair direction. Structures (b) and (d) are equivalent.}
    \label{fig:equivalence}
\end{figure}

Degenerate stacking structures ({\em e.g.}, $(s_1, s_2, \ldots, s_{N-1})$ and $(s_{N-1}, \ldots, s_2, s_1)$) are equivalent since they are related by symmetry ({\bf Figure~\ref{fig:equivalence}}). For example, {\bf Figures~\ref{fig:equivalence}(a)} and {\bf \ref{fig:equivalence}(b)} show two equivalent trilayer structures. Two successive mirror operations are applied to structure (1,2) to establish their equivalence. The first mirror operation, shown in {\bf Figure~\ref{fig:equivalence}(c)}, reverses the monolayer order and stacking arrangement. In this scenario, the bottom and top monolayers have opposite orientations, which requires an additional mirror plane. The second operation, shown in {\bf Figure~\ref{fig:equivalence}(d)}, transforms the structure to match (2,1), demonstrating the equivalence.

The number of nonequivalent structures is thus given by:
\begin{eqnarray}\label{eq:comb_reduced}
   N_{\text{stack}}(D,N)=D^{N-1}-N!
\end{eqnarray}
(We proved {\bf Equation~\eqref{eq:comb_reduced}} for trilayer, tetralayer, and pentalayer stacks by explicit tabulations.)

{\bf Table~\ref{tab:equivalence}} lists the number of non-equivalent stacks up to $N=6$ monolayers as obtained from {\bf Equations~\eqref{eq:comb} and \eqref{eq:comb_reduced}}. {\bf Equation~\eqref{eq:comb_reduced}} calls for significantly fewer options than {\bf Equation~\eqref{eq:comb}}, which facilitates numerical calculations. The nonequivalent structures are shown in bold in {\bf Table~\ref{tab:trilayer}} (and also in {\bf Supplementary Tables S1} and {\bf S2}). The number of nonequivalent structures in this study is smaller than those proposed by Arcudia \textit{et al}. \cite{arcudia2025comprehensive}. This difference results from the exclusion of the local maxima in the energy landscape shown in {\bf Figure~\ref{fig:1}(e)} in this study, which they included.

Intrinsic material properties may differ on equivalent structures. For instance, ferroelectricity may or may not break the equivalence discussed above (see {\bf Supplementary Note S2} and {\bf Supplementary Figure S2}). Therefore, {\bf Equation~\eqref{eq:comb}} is selected to provide a general description of stacking properties in subsequent sections, in the understanding that energies can be readily obtained on the smaller set of structures given by {\bf Equation~\eqref{eq:comb_reduced}}.

\begin{table}[tb]
\caption{\textbf{Metastable stacking configurations without and with energy degeneracies.} The number of stacking configurations for three, four, five, and six monolayers as calculated with {\bf Equations~\eqref{eq:comb} and \eqref{eq:comb_reduced}}.}
\label{tab:equivalence}
\centering
\begin{tabular}{cccc}
\hline
\hline
 &\multicolumn{3}{c}{Counting method} \\
$N$&   $D^{N-1}$   & $D^{N-1}-N!$   \\
\hline
2&4 & 4 \\
3&16&10\\
4&64&40 \\
5&256&136\\
6&1024&304\\
\hline
\hline
\end{tabular}
\end{table}

\subsection*{Distribution of stacking configurations at finite temperature}\label{sec:prob1}

Although the ongoing discussion omits the role of growth substrates, it still permits addressing fundamental questions arising in experimental growth, like the relative likelihood of finding a given configuration on an $N-$monolayer stack. Given the energy of a configuration from {\bf Equation~\eqref{eq:EN}}, Boltzmann weights can be used to estimate such relative probability.

Considering bilayers, the partition function is $Z_2(T)=\sum_{s'_1=1}^{4}\exp(-\mathcal{E}_2^{(s'_1)}/k_B T)$, so that the relative energy of finding the structure $s_1$ is: $\mathcal{P}(s_1)=\exp(-\mathcal{E}_2^{(s_1)}/k_BT)/Z_2(T)$. For trilayers ($N=3$), there are 16 possible energies ({\bf Equation~\eqref{eq:comb}}), and using the energies obtained from {\bf Equation~\eqref{eq:renormalized}} that are listed in {\bf Table~\ref{tab:trilayer}} (or the subset in bold font in the same table, duly accounting for degeneracies), one estimates:
\begin{equation*}
Z_3(T)= \sum_{s'_1,s'_2=1}^4 \exp
\left(
-\frac{\mathcal{E}'_3(s'_1,s'_2)}{k_BT}
\right).
\end{equation*}
Consequently, the probability of finding any of the sixteen possible trilayer structures is:
\begin{equation*}
\mathcal{P}(s_1,s_2)=\text{exp}\left(-\frac{\mathcal{E}'_3(s_1,s_2)}{k_BT}\right)/Z_3(T),
\end{equation*}
with $\mathcal{E}_2^{s_1}$ chosen from $D$ possible energies in {\bf Table \ref{tab:bilayer}} to be assigned to the first interface, and $\mathcal{E}_2^{s_2}$ an independent choice of these $D$ energies corresponding to the second interface. A similar process can be constructed for stacks containing $N$ monolayers with $N-1$ interfaces, and the partition function then becomes:
\begin{equation}
    Z_N(T)=\sum_{s'_1,s'_2\dots,s'_{N-1}=1}^{4}\exp\left(-\frac{\mathcal{E}'_N(s_1,...,s_{N-1})}{k_BT}\right),
\end{equation}
with a probability of finding one of the $4^{N-1}$ possible configurations estimated as:
\begin{equation}\label{eq:probability_N}
\mathcal{P}(s_1,\dots,s_{N-1})=\exp\left(-\frac{\mathcal{E}'_N(s_1,\dots,s_{N-1})}{k_BT}\right)/Z_N(T).
\end{equation}

\begin{figure*}[h!tb]
    \centering
    \includegraphics[width=\textwidth]{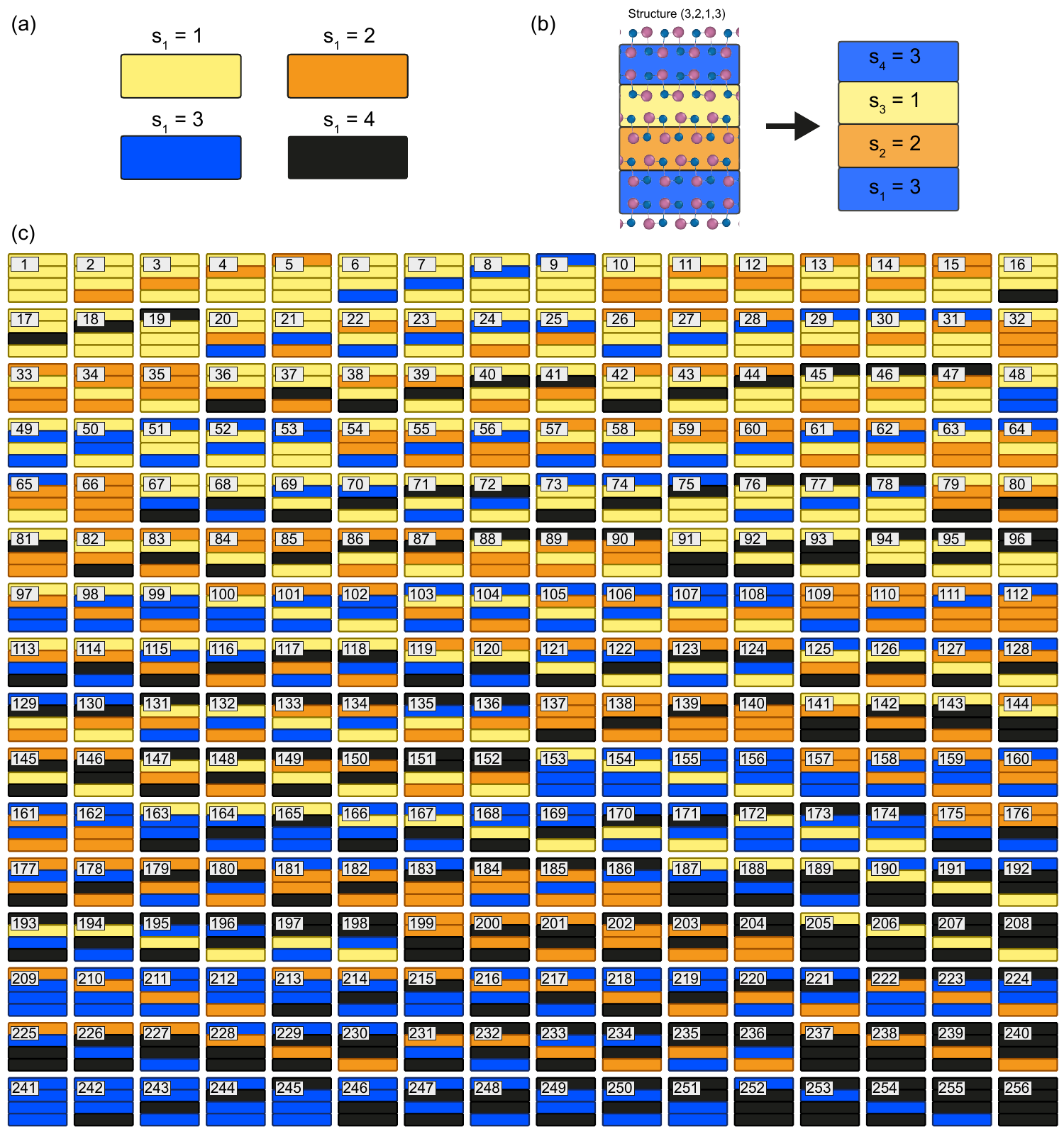}
    \caption{\textbf{Likelihood of finding each of the 256 pentalayer SnSe stacks.} {(a) Colors used to represent the four different bilayer stackings (see {\bf Figure~\ref{fig:1}}). (b) Structure (3,2,1,3). (c) The 256 pentalayer configurations ranked by probability from most to least probable at $T=30$ and 200 K.}}
    \label{fig:possibilities}
\end{figure*}

{Putting everything we have learned to use, the probabilities of finding each pentalayer stacking configuration will be calculated. {\bf Figure~\ref{fig:possibilities}(a)} shows the color convention corresponding to the stacking interface in {\bf Figure~\ref{fig:1}}. We provide the color coding in {\bf Figure~\ref{fig:possibilities}(b)}. All 256 structures, sorted from most to least probable at 30 K and 200 K are shown in {\bf Figure \ref{fig:possibilities}(c)}.}

The most probable structure is the GS, (1,1,1,1), which has a probability of 99.94\% at 30 K, {and 26.46\% at 200 K}. The next most probable structures are all possible permutations of the stack (1,1,1,2): (1,1,2,1), (1,2,1,1), and (2,1,1,1), each with a probability of 0.02\% at 30 K, {and 7.05\% at 200 K}. {This implies that building stacks at higher temperature yields a larger proportion of mismatched, non bulk-like stacks.} A stack with higher energy, with at least one different index, appears after these permutations, followed by all its possible permutations. This pattern is repeated until reaching the least probable structure (4,4,4,4), which is shown in the bottom right of {\bf Figure~\ref{fig:possibilities}}. {\bf Supplementary Table S4} lists the probabilities of occurrence for the fifteen most probable structures at 30 K. Beyond these, probabilities become negligible, with nonzero values appearing only after the eighth decimal.

The nineteen most probable structures for pentalayers (first and second row in {\bf Figure~\ref{fig:possibilities}}) provide insight into the configurations most likely to be observed experimentally when growing few-monolayer SnSe. This likelihood changes due to structural phase transitions, as will be discussed later on this paper.

\subsection*{Optical second harmonic intensities}\label{sec:SHG}

{The SHG patterns of all structures are examined in this section. The aim is to present possible SHG patterns associated with the few-monolayer SnSe samples. This response provides insight into crystal symmetry. The symmetry of the second-order susceptibility tensor ($\chi^{(2)}_{ijk}$), which is required for calculating SHG intensity, is determined by the crystal PG \cite{boyd2008nonlinear}. Stacking configurations within the same PG obey identical symmetry constraints and will show the same symmetry-related angular features in their SHG response. However, the numerical values of their $\chi^{(2)}_{ijk}$ tensor components may differ (see Ref.~\cite{similarSHG}). $\chi^{(2)}_{ijk}$ was calculated using DFT (see {\bf Methods}) for structures with mm2 (bilayer $s_1=2$) and m (stack (1,2)) PGs. These two structures will serve as representatives for all other stacks, since all noncentrosymmetric stacks are classified within these two PGs. A more quantitative determination of stacking dependence would require calculating the frequency-dependent susceptibility for every configuration under identical computational conditions, which is beyond the scope of the present study.}

\begin{figure*}[t!]
    \centering
    \includegraphics[width=\textwidth]{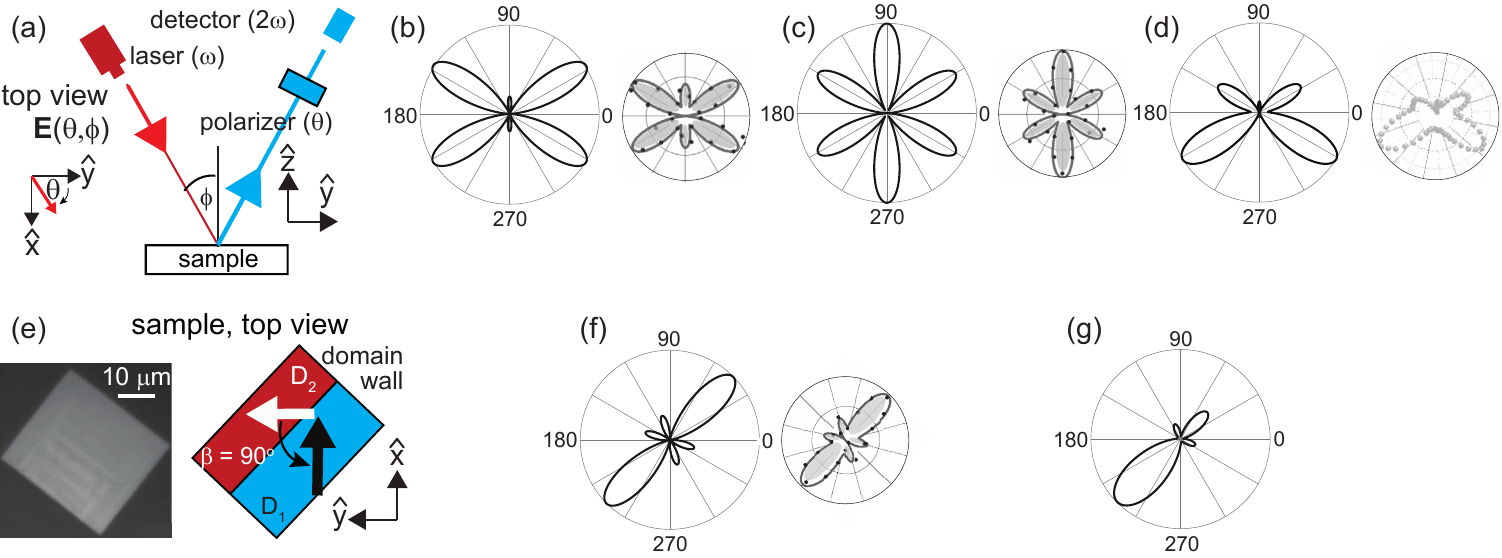}
    \caption{\textbf{SHG characterization of few-monolayer SnSe.} {(a) Illustration of the reflection setup for SHG intensities on a sample subjected to a tilted {\bf E} for $\hbar \omega=1.2$ eV. The SHG signal is collected in the reflected direction after passing through a polarizer parallel to $\bf{E}(\theta,\phi)$. The sample consists of a SnSe stack with either m or mm2 PG symmetry. (b) SHG signal of $s_1=2$ (mm2 PG) compared with measurements (adapted from Figure 3(f) of Ref.~\cite{huang_anisotropy_2025}). (c) SHG signal of stack (1,2) (m PG) compared with SHG measurements adapted from Figure 3(e) of Ref.~\cite{huang_anisotropy_2025}. (d) SHG signal of stack (1,2) under tilted electric field polarization ($\phi=45^\circ$)  compared with experiment adapted from Figure 1(h) of Ref.~\cite{mao2023giant}. (e) Polarized bright-field optical microscopy image adapted from Figure 1(d) of Ref.~\cite{mao2023giant} (left) and schematics of domains (right), D$_2$ is rotated by $\beta=90^\circ$ relative to D$_1$. Arrows denote the armchair direction of each domain. (f) SHG signal obtained from the coherent summation of two domains for in-plane polarization (left) compared with the experimental result (adapted from Figure 2(f) in Ref.~\cite{huang_anisotropy_2025}; right). (g) Coherent SHG signal from two domains under tilted polarization ($\phi=45^\circ$).} }
    \label{fig:SHG1}
\end{figure*}

{The scheme used to obtain the SHG signal is shown in {\bf Figure~\ref{fig:SHG1}(a)}. The laboratory Cartesian frame $(x,y,z)$ is chosen to coincide with that described in {\bf Figure~\ref{fig:1}}. Experiments are typically conducted with the incident light beam perpendicular to the sample, ensuring that the {\bf E} polarization remains within the sample plane \cite{huang_anisotropy_2025,mao2023giant}. Huang \textit{et al.} reported that the stripes observed on their SnSe sample are indicative of height variations attributed to the uneven surface of the cleaved mica substrate used for growth \cite{huang_anisotropy_2025}. Thus, the incident beam may encounter a locally tilted surface, introducing an out-of-plane component to the {\bf E} polarization. For this reason, we consider a laser beam at frequency $\omega$ with an incident angle $\phi$} relative to the surface normal ($z-$axis). The polarization of the incident electric field in the $(x,y,z)$ is $\textbf{E}(\theta,\phi)=E_0(\cos(\phi)\sin(\theta),\cos(\phi)\cos(\theta),\sin(\phi))^T$, where $\theta$ is the polarization angle with respect to the $y$-axis and $E_0$ is the amplitude. The SHG signal at frequency $2\omega$ is analyzed in the reflected direction. A polarizer, aligned parallel to $\textbf{E}(\theta,\phi)$, is placed before the detector to filter the SHG signal. All SHG patterns shown are parallel to $\textbf{E}(\theta,\phi)$. The $\omega$ dependence was omitted to avoid overloading the notation. The theoretical framework used for calculating the SHG intensity parallel to $\textbf{E}(\theta,\phi)$ from $\chi^{(2)}_{ijk}$ is detailed in {\bf Methods} and {\bf Supplementary Note S3}. {The frequency of the incident beam used is 1.2 eV \cite{mao2023giant,huang_anisotropy_2025}}.

{Left panels in {\bf Figures~\ref{fig:SHG1}(b,c,d)} show the calculated SHG from our DFT calculations and {\bf Equations~\eqref{eq:parallel_mm2}}, and {\bf \eqref{eq:parallel_m}}. The right panels in {\bf Figures~\ref{fig:SHG1}(b,c,d)} shows experimental SHG measurements adapted from Refs.~\cite{huang_anisotropy_2025} and \cite{mao2023giant} for comparison.}

{The SHG intensity for the mm2 PG is presented in the left panel of {\bf Figure~\ref{fig:SHG1}(b)}, where the two-fold rotation and the horizontal and vertical mirror planes characteristic of this symmetry are evident. The shape of this SHG signal have been reported experimentally in a SnSe stack, as shown in the right panel of {\bf Figure~\ref{fig:SHG1}(b)} (adapted from Figure 3(f) of Ref.~\cite{huang_anisotropy_2025}).}

{ The SHG intensity for the m PG structure is shown in the left panel of {\bf Figure~\ref{fig:SHG1}(c)}.  Although the m PG breaks two-fold symmetry, this asymmetry is not manifest in the SHG pattern because the in-plane electromagnetic field projects identical $\chi^{(2)}_{ijk}$ in both PGs (see {\bf Supplementary Note S3}).  This shape resembles the experimental measurement presented in the right panel of Figure~\ref{fig:SHG1}(c) (adapted from Figure 3(e) of Ref.~\cite{huang_anisotropy_2025}).  {\bf Figures~\ref{fig:SHG1}(b)} and {\bf Figures~\ref{fig:SHG1}(c)} have identical symmetries (calculated for $\phi=0$).}

{The lobes are no longer symmetric for the structure with m PG when the electromagnetic radiation is tilted ($\phi = 45^\circ$), as shown in the left panel of {\bf Figure ~\ref{fig:SHG1}(c)}, resulting in butterfly-like shapes. Mao \textit{et al.} observed a similar pattern in a SnSe sample, as shown in the right panel of {\bf Figure ~\ref{fig:SHG1}(c)} (adapted from Figure 1(h) of Ref.~\cite{mao2023giant}). Their samples exhibit mixed stacking rather than a single ideal sequence \cite{mao2023giant}, resulting in stacking arrangements with m PG, which coincides with the PG of the (1,2) stack. The $C_2$ breaking observed in the asymmetric lobes in the experimental measure may occur because of variations in surface height \cite{huang_anisotropy_2025}. These variations may locally tilt the incident field relative to the surface normal, allowing $\chi_{ijk}^{(2)}$ entries that do not appear with in-plane polarization of {\bf E}.}

{Tilting the incident field reveals differences between the SHG signals of structures with mm2 and m PGs. This helps distinguish stacks with uniform stacking, which usually have mm2 PG, from those with mixed stacking and lower symmetry, m PG. }



{Multiple SHG contributions from domains with crystallographic orientations differing by $\sim90^\circ$ interfere coherently when the domains are smaller than the laser spot. This interference can enhance or suppress the SHG signal \cite{mao2023giant}. Mao \textit{et al.} observed these domains as bright and dark stripes in polarized bright-field optical microscopy image (left panel of {\bf Figure~\ref{fig:SHG1}(d)}, adapted from Figure 1(d) of Ref.~\cite{mao2023giant}).} To study these patterns, the coherent sum of two SHG signals in a 90$^\circ$ domain is examined. (see right panel in {\bf Figure~\ref{fig:SHG1}(d)}). To facilitate the discussion, it is assumed that the armchair direction of the first domain (D$_1$ in {\bf Figure~\ref{fig:SHG1}(d)}) coincides with the $\hat{x}$ direction of the laboratory coordinate system $(x,y,z)$, while the local crystal orientation $(x’,y’,z’)$ of the second domain (D$_2$ in {\bf Figure~\ref{fig:SHG1}(d)}) is rotated by $\beta=90^\circ$. Details of the theoretical framework of the coherent interference are given in {\bf Methods}.

{The left panel of {\bf Figure~\ref{fig:SHG1}(f)} shows SHG patterns from the coherent superposition of two domains with m PG symmetry misfit by $\beta=90^\circ$ calculated by {\bf Equation~\eqref{eq:coherent}}.  In-plane polarization ($\phi=0$) was used. Two lobes are significantly reduced, and the two-fold symmetry of the PG is clear in the similar amplitude of opposing lobes. A similar pattern was observed experimentally by Huang \textit{et al.} in an SnSe sample (right panel of {\bf Figure~\ref{fig:SHG1}(e)}, adapted from Figure 2(f) of Ref.~\cite{huang_anisotropy_2025}).}

{Figure~\ref{fig:SHG1}(f) displays the signal observed when a tilted electric field is applied to a domain consisting of two structures with m PG and a relative angle of $\pm 90^\circ$. The effect of the domain wall is clear in the reduction of two lobes. The tilting is shown by the asymmetry of the lobes. Most SHG measurements reported in Ref.~\cite{mao2023giant} show asymmetric lobes. This asymmetry may be due to small height variations across the sample. However, none of the reported patterns resembles that of Figure~\ref{fig:SHG1}(f). We expect this pattern may be observed in future experiments.}

{As stated above, few-monolayer SnSe is often grown and characterized within a substrate \cite{huang_anisotropy_2025,sutter_singlecrystalline_2026,wang_heteroepitaxial_2026}, which breaks the out-of-plane inversion symmetry and renders the symmetries listed in {\bf Supplementary Table S3}. As a result, originally centrosymmetric structures may exhibit a non-zero SHG.}

{The patterns presented in this section depict the full range of potential SHG responses in layered SnSe, indicating that the SHG response is not highly sensitive to stacking order. These findings provide valuable insights for interpreting experimental observations.}

\subsection*{Predicting phases at finite temperature}\label{sec:prob2}

{On few-monolayer SnSe stacks, ferroelectric-to-paraelectric phase transitions are not captured by {\bf Equation~\eqref{eq:probability_N}} and another application of this formalism concerns such transitions.}

{We assume the paraelectric phase of these \textit{few-monolayer stacks} to be tetragonal; this is based on experimental results (see Fig.~3(b) of Ref.~\cite{chang2016discovery}, or Fig.~3(a) on Ref.~\cite{sutter2026}) and molecular dynamics evidence (refer to Figs.~1(d) and 1(f) in \cite{mehboudi2016structural} and Fig.~4 of Ref.~\cite{barraza2018tuning}). The most reliable determination of thermal evolution of monolayer group-IV MCs is on Ref.~\cite{chang2016discovery}: there, a SnTe monolayer on a graphitic substrate have equal in-plane lattice parameters at the paraelectric transition, so its lattice is tetragonal. (The ratio between the bulk lattice parameters becomes {\em nearly unity} past $T_c$ even in the bulk; see Figure 1 on Ref.~\cite{chattopadhyay1986neutron}.) \textit{Since we deal with few-monolayer structures here, the assumption of a tetragonal lattice seems granted.} In fact, the motivation for this section is the granularity with which the dependence of structure on temperature has been documented; we hope that this discussion motivates experimentalists to investigate this important matter. }

\begin{figure}[tb]
    \centering
    \includegraphics[width=0.48\textwidth]{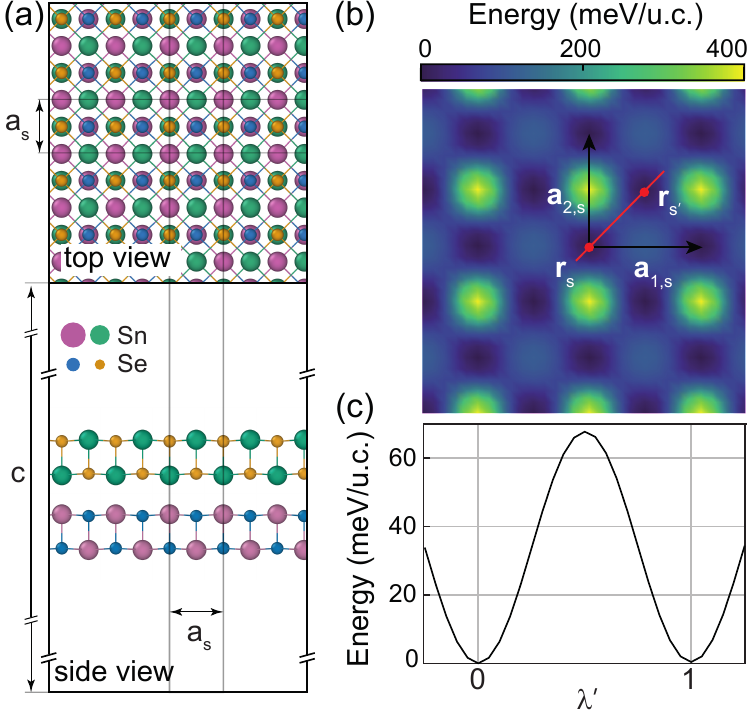}
    \caption{\textbf{Structural and energetic properties of the MC bilayer with a tetragonal lattice.} (a) Top and side views of the tetragonal lattice. (b) The energy landscape of the tetragonal lattice ($\textbf{P}=\textbf{0}$)upon sliding of the upper monolayer relative to the fixed bottom monolayer by $\mathbf{r}$. Minima within the unit cell occur upon  a relative in-plane displacement of the upper monolayer by ${\bf r}_{s}={\bf 0}$ and ${\bf r}_{s'} = {\bf a}_{1,s}/2+{\bf a}_{2,s}/2$. (c) The energy barrier along the path is colored red in panel (b).}
    \label{fig:temp}
\end{figure}

{\bf Figure \ref{fig:temp}(a)} presents the top and side views of the SnSe bilayer in this paraelectric phase. We create an energy landscape to identify the possible energy minima of this phase, which is calculated using the methodology discussed in {\bf Figure~\ref{fig:1}}.

{\bf Figure \ref{fig:temp}(b)} illustrates the energy landscape associated with the in-plane displacement $\mathbf{r}$ of the upper monolayer relative to the bottom one. The energy values are relative to the lowest energy structure. The in-plane lattice vectors are defined as ${\bf a}_{1,s}=(a_s,0)$ and ${\bf a}_{2,s}=(0,a_s)$, where $a_s=4.33$ \AA~\cite{poudel2019group}.

Two minima ($D=2$) are identified within the unit cell: the first, labeled $s$, is located at ${\bf r}_s=0$, and the second, labeled $s'$, is at ${\bf r}_{s'}$. These structures are degenerate and connected by $C_4$ symmetry. However, they are considered inequivalent because, as the temperature approaches the critical transition temperature $T_c$, structures 1 and 3 converge to $s$, while structures 2 and 4 converge to $s'$.

The energy barrier between them is estimated by linearly interpolating the lattice and atomic positions along a continuous path connecting the two degenerate minima at ${\bf r}_s$ and ${\bf r}_{s'}$, as depicted by the red line in {\bf Figure \ref{fig:temp}(b)}. This path is constructed by displacing the upper monolayer along ${\bf r}'(\lambda') = (1-\lambda'){\bf r}_s + \lambda' {\bf r}_{s'}$ for $\lambda' \in [0,1]$. {\bf Figure \ref{fig:temp}(c)} shows the energy profile for this interpolation, which reveals that the two minima are separated by an energy barrier of $\sim$65 meV/u.c. ($\sim$754 K), which prevents phase transitions at room temperature.

Having described the properties of this paraelectric phase, we now initiate a discussion of the structural transition taking place at finite temperature.

\subsection*{Structural phase transition on few-monolayer SnSe stacks at finite temperature}\label{sec:transition}


A ferroelectric-to-paraelectric MC phase transition ensues at finite temperature. As far as we know, there is no theory studying such behavior on few-monolayer stacks yet. Representing the most speculative part of this article, a discussion is started in that direction. To this end, the order parameter $\Delta\alpha_0=\Delta \alpha(T=0)$ \cite{chang2016discovery} is written down \cite{barraza2018tuning}:
\begin{equation}\label{eq:order}
   \frac{a}{b}=\frac{1+\sin(\Delta\alpha)}{\cos(\Delta\alpha)}\simeq 1+\Delta\alpha\text{ when $\Delta\alpha\simeq$ 0},
\end{equation}
($\Delta\alpha\simeq 0$ here) where $a$ and $b$ are the in-plane lattice parameters at $T=0$, defined in {\bf Figure~\ref{fig:1}} with $a>b$. $T_c$ is the temperature at which the lattice adopts a centrosymmetric configuration.

{$\Delta\alpha(T)$ takes finite values when $T<T_c$ in the ferroelectric phase (orthorhombic lattice), and it becomes zero when $T>T_c$ as the structure becomes centrosymmetric \cite{chang2016discovery}:}
\begin{equation}\label{eq:alpha}
\Delta \alpha (T)= \Delta \alpha_0\left(1-\frac{T}{T_c}\right)^{1/3}.
\end{equation}

{$\Delta\alpha(T)$ ranges from a maximum value in the ferroelectric phase at $T=0$ K onto zero in the paraelectric phase above $T_c$. }


Emulating ``finite temperature'' monolayer structures {(see Ref.~\cite{barraza2018tuning})}, the atomic positions and lattice vectors corresponding to each minimum $s_1$ are linearly interpolated toward those of the tetragonal lattice. The interpolated lattice vectors for each structure are expressed as $\bar{{\bf a}}_1(\kappa) = (1-\kappa){\bf a}_{1,s} + \kappa {\bf a}_{1}$ and $\bar{{\bf a}}_2(\kappa) = (1-\kappa){\bf a}_{2,s} + \kappa {\bf a}_{2}$, where $\kappa \in [0,1]$. Atomic coordinates are similarly  interpolated using $\bar{\bf x}_i(\kappa) = (1-\kappa){\bf x}_{s_1,i} + \kappa {\bf x}_{sq,i}$, where ${\bf x}_{s_1,i}$ denotes atom $i$ on a bilayer ($s_1 \in \{1,2,3,4\}$), ${\bf x}_{sq,i}$ is the atomic position in one of the tetragonal structures ($s$ or $s'$), and $i$ runs from 1 to 8.


{{\bf Figure~\ref{fig:probs}(a)} shows the DFT total energies of a SnSe monolayer along these paths. The solid, filled (red) triangles denote the same energy, at which the four structures turn degenerate. The order parameter at each interpolation point was determined according to {\bf Equation~\eqref{eq:order}}.} {These trends are modeled by a Fermi–Dirac-type function for $0=\Delta\alpha(T_c)<\Delta\alpha<\Delta\alpha(0)$ for convenience}:
\begin{equation}\label{eq:extrapolation_alpha}
   \mathcal{E}_2^{(s_1)}(\Delta \alpha(T))= \mathcal{E}_0+\frac{\bar{\mathcal{E}}^{(s_1)}_2}{1+\exp\left(-\frac{(\Delta\alpha(T)-\Delta \bar{\alpha})}{\delta}\right)},
\end{equation}
{where $\mathcal{E}_0,$  $\bar{\mathcal{E}}^{(s_1)}_2$, $\Delta \bar{\alpha},$ and $\delta$ are free parameters.}
{The energy of each structure $s_1\in \{1,2,3,4\}$ yields $\mathcal{E}_0=\{-24.1291\pm0.0003,-24.1297\pm0.0002,-24.1285\pm0.0005,-24.1299\pm0.0001\}$ meV, $\bar{\mathcal{E}}^{(s_1)}_2=\{-44.0\pm0.8,-22.9\pm0.5,-64.8\pm 1.4,-18.3\pm 0.3\}$ $\mu$eV, $\Delta \bar{\alpha}=\{28.9\pm 0.3,17.9\pm 0.2,37.3\pm 0.4,18.1\pm 0.2\} \times 10^{-3}$, and $\delta=\{8.1\pm 0.3,4.6\pm 0.2,9.4\pm 0.4,5.2\pm 0.2\} \times 10^{-3}$ for structures 1, 2, 3, and 4, respectively. Equation \eqref{eq:extrapolation_alpha} is thus a continuous expression on the energy that can be attached to Boltzmann-like thermal estimates straightforwardly:} {that equation represents the initial stage in formulating the thermal dependence of the structural energy as the structures turn from ferroelectric or antiferroelectric onto a paraelectric one.}


The thermal dependence of the energy on $\Delta\alpha$ is examined by substituting {\bf Equation~\eqref{eq:alpha}} {(which is firmly grounded in both experiment \cite{chang2016discovery} and theory \cite{barraza2018tuning})} into {\bf Equation~\eqref{eq:extrapolation_alpha}}:
\begin{align}\label{eq:E_T}
\mathcal{E}_2^{(s_1)}(T)&=\mathcal{E}_2^{(s_1)}\Big[1+\notag\\
&\quad\exp\Big(\frac{1}{\delta}\Big[\Delta\bar{\alpha}-\Delta\alpha_0\left(1-\frac{T}{T_c}\right)^{1/3}\Big]\Big)\Big]^{-1}.
\end{align}
{{\bf Equation~\eqref{eq:E_T}} incorporates the ferroelectric-to-paraelectric transition into the energies of all bilayer structures.}



\subsection*{Critical temperature in few-monolayer SnSe}
{Applying Equation \eqref{eq:E_T} still requires setting $T_c$ for few-monolayer SnSe explicitly. But only a handful of $T_c$ values have been reported \cite{chang2020microscopic,mehboudi2016structural,sutter,sutter2026,chattopadhyay1986neutron} . We estimate $T_c$ from existing reports. The model thus built is physically grounded on (a) tetragonal paraelectric few-monolayers; (b) a well-known structural path between ferroelectric and paraelectric monolayers; and (c) available $T_c$ data.}

{Note that, when increasing the number of monolayers, $T_c$ increases {\em to saturate in the bulk}: Indeed, take SnSe, for which the {\em monolayer} paraelectric phase takes hold at 400 K when grown on a graphitic substrate \cite{chang2020microscopic}; for which Sutter {\em et al.} observe a $T_c$ of about 623 K for thicknesses ranging between 10 and 20 nm, roughly corresponding to about 20 to 40 monolayers \cite{sutter2026} and 593 K for 14 to 36 monolayers \cite{sutter}; and for which the bulk phase transitions at 887 K \cite{chattopadhyay1986neutron}. That data is shown by orange circles on {\bf Figure \ref{fig:probs}(b)}.}

{Additionally, substrate-free \textit{ab-initio} molecular dynamics calculations of SnSe monolayers (with $T_c \sim 175$ K \cite{mehboudi2016structural} and 212 K \cite{barraza2018tuning}) and bilayers ($T_c \sim 225$ K) also show an increase on $T_c$ as the number of monolayer increases. Fluctuations on estimated $T_c$ from molecular dynamics are due to the effect of exchange-correlation approximations and system size. The granularity of the data precludes, at this stage, to be too detailed and include changes on $T_c$ on a given multilayer system due to stacking; Sutter {\em et al.} do indicate that to be the case \cite{sutter2026} though.}

{To give a sensible range of $T_c$, we produced two curves on {\bf Figure \ref{fig:probs}(b)}: One of these takes all the available experimental data on $T_c$ for SnSe; the other takes the calculated $T_c$ for monolayers and bilayers, and the experimental data on thicker samples, to which simulation should converge to. We explicitly observe that substrates should increase $T_c$ on grown films, so that the increase on $T_c$ due to a substrate is indeed physically sensible. We use an exponential function (which we also deem to be physically sensible) to describe the monotonic, saturating trend:}
\begin{equation}\label{eq:fitting}
    1/N = C \exp(-\gamma T_c).
\end{equation}
{This yields lower (dashed curve) and upper (solid curve) bounds for $1/N$ \textit{versus} $T_c$ on {\bf Figure~\ref{fig:probs}(b).} The excursion on $T_c$ in between the two curves reflects the uncertainty and granularity of the available data. The solid curve has $C = 11.38 \pm 7.18$ and $\gamma=0.019 \pm 0.001$ 1/K. The dashed curve uses $C=1977.8 \pm 1633.2$ and $\gamma = 0.019 \pm 0.001$ 1/K.} {The uncertainty on $T_c$ in \textbf{Figure~\ref{fig:probs}(b)} permeates all subsequent results, which are thus reported within lower and upper $T_c$ bounds.} {The structural energies and the the probability of finding a specific configuration for arbitrary $N$ at finite $T$ is analyzed using the established range of $T_c$.}

\subsection*{Thermodynamic likelihood of stacking configurations}\label{sec:prob3}

\begin{figure*}[t]
    \centering
    \includegraphics[width=\textwidth]{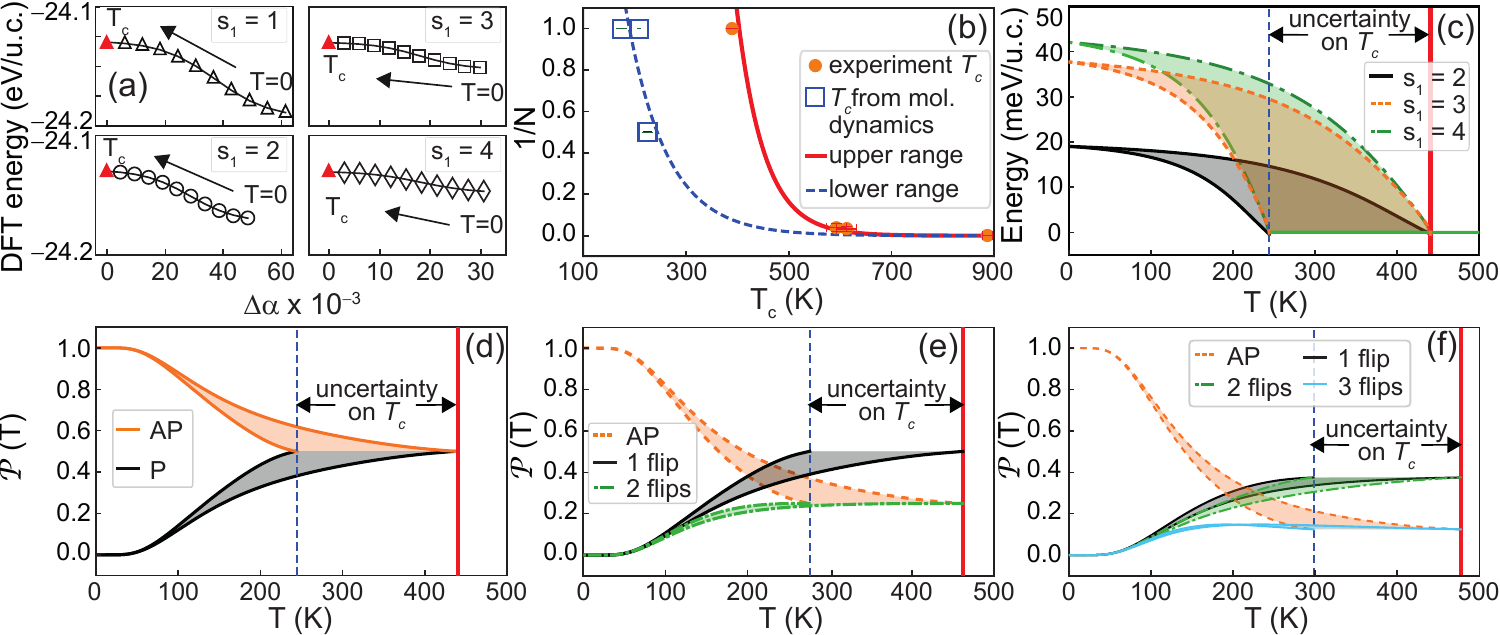}
    \caption{\textbf{Temperature-driven structural transitions and stacking probabilities for bilayer, trilayer, and tetralayer SnSe.} {(a) Bilayer DFT energy along the structural interpolation path between the orthorhombic and tetragonal lattices. The red filled symbol are the points where all structures are energy-degenerate. (b) $1/N$ \textit{versus} $T_c$ for an arbitrary number of monolayers. Orange dots and blue squares are the experimental and theoretical data from previous reports \cite{chang2020microscopic,mehboudi2016structural,mehboudi2016structural,sutter,chattopadhyay1986neutron,sutter2026}. The upper (red line) and lower (dashed blue line) bounds define the uncertainty of $T_c$ for each number of monolayers given the limited available published data. (c) Temperature-dependent energy profiles of bilayers using the upper and lower range shown in panel (b), relative to that of the $s_1=1$ stack for a finite temperature, along the ferroelectric-to-paraelectric transition. (d) Probability of finding an AP or P configuration in bilayers. The AP stacking is preferred below $T_c$. (e,f) Probability of finding structures with monolayer rotations relative to the ground-state configuration in (e) trilayers and (f) tetralayers. The shaded regions in panels (c–f) indicate the uncertainty arising from the uncertainty in $T_c$.}}
    \label{fig:probs}
\end{figure*}

{{\bf Figure~\ref{fig:probs}(c)} shows energies obtained from {\bf Equation~\ref{eq:E_T}} using the upper and lower $T_c$ values from {\bf Figure~\ref{fig:probs}(b)}. The $T_c$ limits appear as vertical red line and a blue dashed line. Energies are relative to $s_1=1$, with uncertainty shown by the shaded region. The temperature where all structures become degenerate now has the uncertainty that comes from $T_c$.}

{\bf Figure~\ref{fig:probs}(d)} presents the probability of finding P and AP bilayers at finite temperature, obtained by substituting {\bf Equation~\eqref{eq:E_T}} into {\bf Equation~\eqref{eq:probability_N}}. Centrosymmetric structures are favored at all temperatures below {a range of $\sim$244 K and $\sim$440 K} (bilayer $T_c$), consistent with the GS identified by DFT calculations. Both probability curves converge to 0.5 {in the $T_c$ range}, which indicates that all structures become tetragonal.

To study the probability of finding specific structures in trilayers and tetralayers at finite temperature, {\bf Equation~\eqref{eq:E_T}} is substituted into {\bf Equation~\eqref{eq:renormalized}}, and the result is then substituted into {\bf Equation~\eqref{eq:probability_N}}. {\bf Figure \ref{fig:probs}(e)} presents the probabilities of finding AP structures and structures with one and two flips (see {\bf Supplementary Table S5} for details of the structures) in trilayers. AP structures are predominant below {a $T_c$ range of $\sim$ 200 K and $\sim275$ K}, while structures containing at least one flip become more probable at higher temperatures.

The tetralayer structures can be classified within four categories: AP structures, and structures with one, two, or three flips (see {\bf Supplementary Table S5} for details of the structures). The probabilities of finding metastable structures below the estimated $T_c$ are shown in {\bf Figure~\ref{fig:probs}(f)}. At low temperatures, AP structures are favored. {Between 200 K $\simeq T_c \simeq$ 250 K}, the preference shifts toward structures with one or two flips. Structures with the largest number of flips are the least probable. Those with few flips always predominate at high temperatures.

{This is the insight of this section: (1) The AP phase becomes the most probable one as monolayers are added because of an increased Boltzmann weight; it becomes the preferred one in the bulk and it has a well-defined $T_c$. (2) The uncertainty on $T_c$ goes beyond a particular stacking sequence: experimentally, it originates from \textit{lack of exact knowledge} of (i) the number of monolayers $N$, (ii) the stacking sequence and, for thin stacks, (iii) the interaction with a substrate. Theoretically, $T_c$ also depends (iv) on exchange-correlation functionals and (v) on the strength of vdW interactions. (3) The finite-temperature model shows that during sample growth, the material may not reach the ground state but become trapped in a metastable configuration. The growth procedure for SnSe monolayers typically involves heating the material to elevated temperatures, followed by cooling. As the temperature decreases below $T_c$, the structure adopts a stacking arrangement determined by the available metastable minima. Results for trilayers and tetralayers ({\bf Figure~\ref{fig:probs}(e,f)}) indicate that configurations with flipped monolayers are more likely to form than AP (bulk-like) structures. This tendency is commonly observed in SnSe samples \cite{sutter,mao2023giant}.}

{In summary, a general framework was developed to enumerate and characterize metastable stacking configurations in few-monolayer MCs, beginning with the local minima identified in bilayers. The energy of these configurations can be accurately described as a sum of bilayer contributions. }{We also show that most probable stacking configurations belong to the mm2 or m PGs. This classification enables the categorization of SHG patterns of metastable stacks into two families. The lower symmetry associated with m PG is observed in SHG patterns only when the incident electric field is tilted out of plane. Additionally, possible SHG patterns resulting from the coherent superposition of signals from different domains were presented.}

We also incorporated thermal effects into the additive energy framework to determine the probability of metastable stack formation at finite temperatures.

This microscopic perspective explains the observation of ferroelectricity in experimental samples, even though the GS is not expected to exhibit it.

\section*{Methods}\label{sec:methods}
\subsection*{Details of DFT calculations}
DFT calculations of the structure of ultrathin SnSe were performed with the VASP package \cite{kresse1996efficiency,kresse1996efficient2}. We used PAW pseudopotentials \cite{PAW}, a Monkhorst-Pack 12$\times$12$\times$1 $k$-point mesh, an energy cutoff of 600 eV, electronic energy convergence criteria of $10^{-8}$ eV, and a force convergence of $10^{-6} ~\text{eV/\AA}$. We added 40 \AA ~ of vacuum along the $z-$direction to reduce self-interactions among periodic copies. The van der Waals interaction was included within the optPBE-vdW functional \cite{optPBE}. Phonon calculations were performed using a 5$\times$7$\times$1 for $s_1=1$ and 5$\times$5$\times$1 for the other structures. A 3$\times$3$\times$1 $k$-point mesh was used. Energy landscapes were created by sliding one monolayer away from a minimum-energy configuration without further optimization. The local minima were identified and fully relaxed with respect to both lattice vectors and ionic coordinates. ${\bf P}$ was calculated with the Berry phase formalism \cite{Resta} in select structures. {The second-order susceptibility tensor was calculated using the DFT-D2 vdW correction in the TINIBA code  \cite{cabellos_effects_2009}.}

\subsection*{SHG signal calculation}

Within the $(x,y,z)$ coordinate system, the components of the second-order nonlinear polarization (${\bf P}^{(2)}$) induced by the incident light for a SnSe stack are expressed as follows:
\begin{equation}\label{eq:SOP}
P^{(2)}_i(\theta,\phi) = \sum_{jk} \chi_{ijk}^{(2)} E_j(\theta,\phi) E_k(\theta,\phi),
\end{equation}
where $E_{j}(\theta,\phi)$ and $E_{k}(\theta,\phi)$ denote the components of $\bf{E}$. Using {\bf Equation~\eqref{eq:SOP}}, the SHG intensities, parallel to $\textbf{E}(\theta,\phi)$ for mm2 and m PGs, are given by:
\begin{eqnarray}\label{eq:paralell}
I_{||,\text{mm2}}&=&\frac{1}{E_0^3}|\textbf{P}^{(2)}_{\text{mm2}}(\theta,\phi)\cdot \textbf{E}(\theta,\phi)|^2,\label{eq:parallel_mm2}\\
I_{||,\text{m}}&=&\frac{1}{E_0^3}|\textbf{P}^{(2)}_{\text{m}}(\theta,\phi)\cdot \textbf{E}(\theta,\phi)|^2,\label{eq:parallel_m}
\end{eqnarray}
where $\textbf{P}^{(2)}_{\text{mm2}}(\theta,\phi)$ and $\textbf{P}^{(2)}_{\text{m}}(\theta,\phi)$ are the polarization of the mm2 and m PGs, respectively. The explicit forms are provided in {\bf Supplementary Note S3}.

The superposition of the nonlinear polarization of two adjacent ferroelectric SnSe domains is examined. The polarization of domain D$_1$ is given in {\bf Equation~\eqref{eq:SOP}}. For domain D$_2$, it is first expressed in $(x',y',z')$ and subsequently transformed to $(x,y,z)$

The electric field in the $(x’,y’,z’)$ system is expressed as:
\begin{eqnarray}\label{eq:Erotation}
\bar{\mathbf{E}}(\theta,\phi,\beta)=R^z(-\beta)\mathbf{E}(\theta,\phi),
\end{eqnarray}
where $R^z$ denotes the rotation matrix about the $z$-axis. In this coordinate system, the nonlinear polarization is expressed as:
\begin{eqnarray}\label{eq:Pcrystal}
\bar{P}_{i}^{(2)}(\theta,\beta) = \sum_{jk} \chi_{ijk}^{(2)}\bar{E}_{j}(\theta,\phi,\beta) \bar{E}_{k}(\theta,\phi.\beta),
\end{eqnarray}
where $j$ and $k$ run over the Cartesian indices. This equation can be reformulated in terms of the electric fields in the $(x,y,z)$ system by substituting {\bf Equation~\eqref{eq:Erotation}} into {\bf Equation~\eqref{eq:Pcrystal}}:
\begin{align}
\bar{P}_{i}^{(2)}(\theta,\phi,\beta) &= \varepsilon_0 \sum_{jk} \sum_{lm}\chi_{ijk}^{(2)} \nonumber\\
&\times R^{z}_{jl}(-\beta)R^{z}_{km}(-\beta) E_l(\theta,\phi) E_m(\theta,\phi).
\end{align}
where $l$ and $m$ are Cartesian indices.

The second-order polarization is transformed back to the $(x,y,z)$ coordinate system using the relation $P’^{(2)}_i(\theta,\phi,\beta)=R^z(\beta)\bar{P}_{i}^{(2)}(\theta,\phi,\beta)$. The total polarization components resulting from the coherent summation of the two domain walls in the laboratory coordinate system are written as:
\begin{equation} \label{eq:P}
P^{(2)}_{\text{tot},i}(\theta,\phi,\beta) =P^{(2)}_i(\theta,\phi)+P’^{(2)}_i(\theta,\phi,\beta).
\end{equation}
Then, the parallel component of the total polarization intensity is:
\begin{equation}\label{eq:coherent}
I’_{||}
=\frac{1}{E_0^3}|\mathbf{P}^{(2)}_{\text{tot}}(\theta,\phi,\beta)\cdot \mathbf{E}(\theta,\phi,\beta)|.
\end{equation}
The polarizations of the mm2 and m PGs shown in {\bf Supplementary Note S3} can be replaced in {\bf Equation~\eqref{eq:coherent}} to obtain the SHG signal. The $\chi_{ijk}^{(2)}$ for both PGS were calculated using the TINIBA code \cite{cabellos_effects_2009}. Excitonic effects were not considered.

\section*{Data Availability}
The VASP input files used in this study are available on Figshare at https://doi.org/10.6084/m9.figshare.33235458. Python scripts and SHG data are available from the authors upon request.

\section*{Author contributions}
S.B.L.~designed the research. J.D.M.~developed models and symmetry analyses and performed DFT calculations, with the exception of those of second-order optical susceptibilities. S.B.L. and J.D.M. determined the combinatoric formula derived on this paper. J.L.C. calculated second-order optical susceptibilities on TINIBA, a code developed by B.S.M. S.B.L. and J.D.M. wrote the manuscript.

\section*{Competing interests}
The authors declare no competing financial or non-financial interests.

\section*{Acknowledgements}
J.D.M.~and S.B.L.~acknowledge financial support from the NSF's Q-AMASE-i program (Award DMR-1906383). Calculations were performed at the Arkansas High Performance Computing Center (NSF Award OAC-2346752). We acknowledge conversations with Hiroyuki Nakamura and Pradeep Kumar.

%


\end{document}